\documentclass[figures,cite,bm]{epl}
\usepackage{amsfonts,amssymb,amsmath}

\title{Enhanced cooperativity below the caging temperature of glass-forming liquids} %
\shorttitle{Enhanced cooperativity in glass-forming liquids} %

\author{B. M. Erwin\inst{1} \and R. H. Colby\inst{1}\thanks{E-mail: \email{rhc@plmsc.psu.edu}} \and S. Y. Kamath\inst{1} \and S. K. Kumar\inst{2}} %
\shortauthor{B. M. Erwin\etal} %
 \institute{ %
  \inst{1} Department of Materials Science and Engineering and the Materials Research Institute
  The Pennsylvania State University - University Park, PA 16802, USA\\
  \inst{2} Department of Chemical and Biological Engineering,
  Rensselaer Polytechnic Institute - Troy, NY 12180, USA} %


\begin{document}

\maketitle

\begin{abstract}
The utility of a cooperative length scale for describing the
dynamics of small molecule glass-formers is shown. Molecular
Dynamics and Monte Carlo simulations reveal a distribution of
cooperatively moving fractal events below the temperature $T_A$ at
which dynamics become caged. Guided by these results, four
straightforward methods emerge to recognize $T_A$ in experimental
data and quantify the length scale that grows on cooling below
$T_A$. This length scale is consistent with 4-D NMR experiments
which are sensitive to the slow moving population.
\end{abstract}

Many liquids either cannot crystallize or crystallize sufficiently
slowly that they vitrify below their glass transition temperature
$T_g$. A fundamental understanding of glass formation is still
lacking because it has not been firmly established whether the
pronounced slowing down is simply kinetic in origin or there is an
underlying thermodynamic character \cite{Adam1965, Kob1999,
Angell2000A, Sillescu1999, Ediger2000, Donth2002A}. This letter
shows that there is a natural length scale for cooperative motion
that grows as the glass transition is approached.

The 1965 model of Adam and Gibbs \cite{Adam1965} suggests that there
should be cooperative motion in glass-forming liquids. The size
$\xi$ of the cooperative volume is related to the configurational
entropy of the liquid $S_c$. At temperatures $T$, sufficiently above
$T_g$, where the relaxation time $\tau_{\alpha}$, and the viscosity
$\eta$, vary as $\tau_\alpha \sim \eta/T \sim \exp\left[ E/k_BT
\right]$, all molecules undergo independent local Brownian movements
without signs of cooperativity. As temperature is lowered, the
density of the liquid gradually increases and Brownian motion
becomes hindered, as neighboring particles block each others
attempts to move. This crowding leads to \emph{cooperative dynamics}
\cite{Kob1999, Angell2000A, Kisliuk2000, Glotzer2000, Binder2003},
active for all $T$ below the \emph{caging temperature} $T_A$. The
onset of cooperativity is also accompanied by the observed `caging
effect' in the mean-square displacement of a particle between the
ballistic and self-diffusive regimes, and a reduction in $S_c$,
causing $\xi$ to grow. These changes result in a progressively
stronger temperature dependence of $\eta$ and $\tau_{\alpha}$ at
lower temperatures.

Since experimental attempts to identify the length scale for
cooperative motion have met with limited success \cite{Sillescu1999,
Ediger2000, Donth2002A}, the dominant evidence for this quantity is
from computer simulations \cite{Glotzer2000, Binder2003,
Mountain1998, Donati1998, Donati1999A, Glotzer1999b, Muranaka2000b,
Kamath2002, Stanley2003, Lacevic2002}. Simulations have the profound
advantages of direct observation of motion and straightforward
identification of both the \emph{size and shape} of cooperatively
rearranged regions. Equilibrium simulations of liquids are not yet
possible near $T_g$, but have been done down to $0.7T_A$
\cite{Kamath2002}. In addition to confirming the essential aspects
of the Adam and Gibbs model, simulations have provided two novel
insights. Instead of a single size scale for cooperative motion,
there is in fact a \emph{broad distribution of size scales} below
$T_A$ \cite{Glotzer2000, Mel'cuk1995, Mountain1998, Donati1998,
Donati1999A, Glotzer1999b, Muranaka2000b, Kamath2003, Stanley2003}.
The largest size  in this distribution $\xi$ grows rapidly as
temperature is lowered, as expected by Adam and Gibbs
\cite{Adam1965}. The second important observation is that the
cooperatively rearranging regions are not the three-dimensional
volumes that were initially proposed, but instead are \emph{fractal}
\cite{Glotzer2000, Donati1998, Donati1999A, Muranaka2000b,
Kamath2003}. The observed fractal dimension (of order $2$) clearly
shows that the majority of molecules within the volume $\xi^3$ have
\emph{not} participated in the cooperative motion. Consequently, a
new model for cooperative motion was proposed that accommodates
these new insights \cite{Colby2000}. All glass-forming liquids show
a temperature dependence of cooperative size scale, with
commensurate effects in viscosity and relaxation time, in reasonable
accord with the expectation of dynamic scaling \cite{Colby2000,
Erwin2002} in the temperature range $T_C<T<T_A$.
\begin{equation}\label{ksi}
\xi^6 \sim \tau_\alpha e^{-E_\alpha/k_BT} \sim \eta e^{-E_\eta/k_BT}
\sim (T-T_C)^{-9}
\end{equation}
The critical temperature $T_C$, is slightly \emph{below} $T_g$ and
is the temperature at which the \emph{equilibrium} extrapolated
values of $\xi$, $\tau_\alpha$ and $\eta$ all diverge. The
material-specific activation energy apparently depends on whether
segmental relaxation ($E_\alpha$) or viscosity ($E_\eta$) is
measured, with $E_{\eta}/E_{\alpha}\simeq 1.2$ for non-polymeric
organic glasses.

\begin{figure}[t]
\centerline{\includegraphics[width=3.1in,keepaspectratio=true]{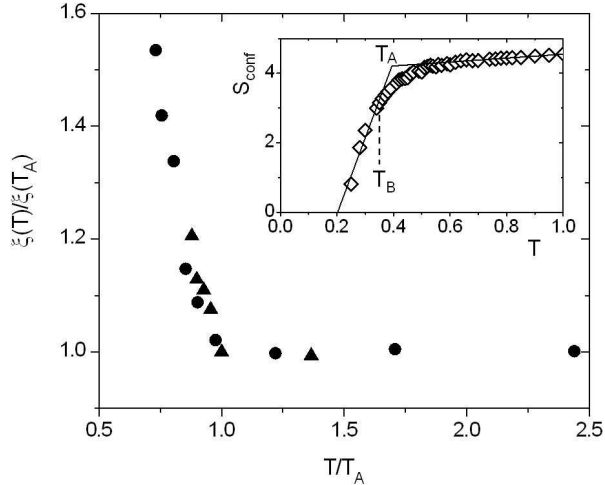}}
\caption{\label{simulation} Simulation results for the length scale
of cooperative motion as a function of reduced temperature. From
molecular dynamics simulations of Lennard-Jones sphere mixtures
($\blacktriangle$ \cite{Lacevic2002}) and from Monte Carlo
simulations of the bond fluctuation model at density $0.8$ for
chains of $10$ monomers ($\bullet$ \cite{Kamath2003}). Inset is the
temperature dependence of the configurational entropy for chains of
10 monomers ($\diamond$ \cite{Kamath2002}).}
\end{figure}

Figure~\ref{simulation} shows data for $\xi$ as obtained from
Molecular Dynamics simulations \cite{Glotzer2000} and new results
from Monte Carlo (MC) simulations on the bond fluctuation model
\cite{Kamath2002}.
%---------- 1st ref to Simulation -
%----------------------------------
In the MC simulations the mobile particles are identified as those
that move over a time scale of interest. We then look for clusters
of these mobile particles and find that their sizes are a function
of time for short times, but quickly become time independent up to
$\tau_\alpha$. We only consider cluster sizes in this intermediate
time range. Figure \ref{simulation} shows mean sizes of the mobile
particle clusters as a function of $T$. Both sets of simulations
show that $\xi$ is sensibly independent of temperature above $T_A$,
but then grows rapidly when temperature is lowered below $T_A$. The
very different nature of the simulations used for the data in
Fig.~\ref{simulation} strongly points to the existence of a growing
length scale below at $T<T_A$. The inset of Fig.~\ref{simulation}
shows that $S_c$ as defined by Adam and Gibbs also changes character
at $T_A$ with a broad crossover between $T_B$ \cite{Stickel1996}
below which Eq.~\ref{ksi} or Vogel-Fulcher should describe dynamics
and a much higher temperature above which dynamics are Arrhenius
\cite{Stickel1996}.

Guided by simulations, and owing to the abrupt change in the very
nature of relaxation at $T_A$ \cite{Binder1996, Baschnagel1997,
Zhang2000, Kamath2002, Binder2003}, the caging temperature is easily
identified by a variety of experiments probing liquid dynamics. We
demonstrate this point with four dynamics experiments that have
broad dynamic range: rotation and translational diffusion of
molecular probes, self-diffusion, dielectric spectroscopy and
rheology. Molecular probe and self-diffusion techniques are
particulary noteworthy because they provide a \emph{model
independent} measure of $T_A$. The length scales extracted from
these experiments, which are in \emph{quantitative} agreement with
existing $\xi(T)$ data from 4-D NMR \cite{Reinsberg2002, Qiu2002},
show a strong temperature dependence \emph{only below} $T_A$.

\begin{figure}[t]
\centerline{\includegraphics[width=3.1in,keepaspectratio=true]{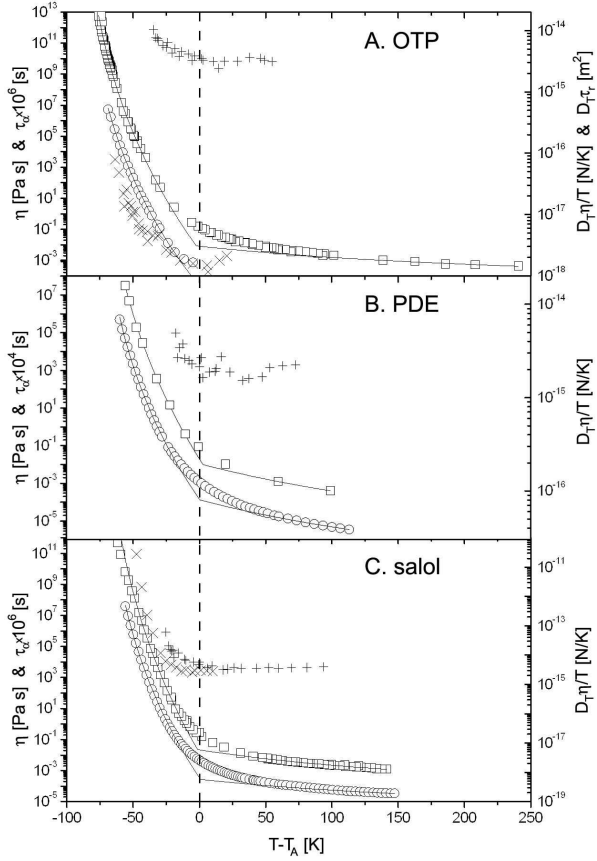}}
\caption{\label{TaPLOT} Dynamic data for OTP, PDE and salol above
and below $T_A$. Temperature dependence of viscosity ($\Box$),
$\tau_{\alpha}$ ($\circ$), probe diffusion ($\times$) and
self-diffusion ($+$). Solid curves are fits of Eq.~\ref{ksi} to data
below $T_A$ and Arrhenius fits to data above $T_A$ for both
viscosity and dielectric data. The dashed line is $T_A$ taken from
the extrapolation of dynamic scaling to the Arrhenius temperature
dependence of viscosity from high-$T$. See Table~\ref{Table} for
references.}
\end{figure}

$T_C$ is determined by using literature data on $\tau_\alpha$ and
$\eta$ well below $T_A$ for glycerol, \emph{o}-terphenyl (OTP),
phenolphthaleinedimethylether (PDE), salol and {\scriptsize\sf
D}-sorbitol and fitting them to Eq.~\ref{ksi} (Table~\ref{Table}).
%---------- 1st ref to Table -
\begin{table}[t]
\begin{center}
\caption{Temperatures and length scales for five glass-forming
liquids. All $T_g$ are from B\"ohmer, et al.~\cite{Bohmer1993A} with
the exception PDE where $T_g=T(\tau_\alpha=100\un{sec})$. All $T_B$
are the crossover temperature between Vogel-Fulcher fits with the
exception of {\scriptsize\sf D}-sorbitol \cite{Wagner1998} and
glycerol \cite{Donth2002A} from $\alpha\beta$-merging.}
\label{Table}
\begin{largetabular}{lllllllll}
Property & $T_A$ [K] & $T_B$ [K] & $T_g$ [K] & $T_C$ [K] & $r_{vdW}$
[\AA] & $\xi(T_A)$ [\AA] & $\xi(T_g)$ [\AA] & References\\ \hline %
PDE & $355$ & $325$ & $294$  & $278$ & $4.1$ & & &
\cite{Stickel1996, Chang1997, Heuberger1996, Stickel1995}\\
{\scriptsize\sf D}-sorbitol & $350$ & $335$ & $274$ & $257$ & $3.4$
& $3.8\pm 1.8$ & $27.\pm 13.$ & \cite{Nakheli1999, Nozaki1998,
Naoki1993A}\\
OTP & $315$ & $290$ & $241$ & $227$ & $3.7$ & $3.5\pm 1.0$ & $52.\pm
18.$ & \cite{Chang1997, Stickel1995, Cicerone1995A, Cukierman1973,
Plazek1994, Laughlin1972, Fujara1992, Greet1967}\\
salol & $275$ & $265$ & $218$ & $204$ & $3.5$ & & &
\cite{Stickel1996, Chang1997, Heuberger1996, Stickel1995}\\
glycerol & $280$ & $262$ & $190$  & $173$ & $2.7$ & $1.9\pm 0.8$ &
$25.\pm 11.$ & \cite{Chang1997, Stickel1995, Schroter2000,
Cukierman1973, Laughlin1972}
\end{largetabular}
\end{center}
\end{table}
%-----------------------------
$T_A$ is defined from the crossovers in Fig.~\ref{TaPLOT}. Different
dynamic experiments provide consistent determinations of both $T_A$
and $T_C$. Table \ref{Table} also reports the crossover temperature
$T_B$ from Donth \cite{Donth2002A} where dynamics switch from one
Vogel-Fulcher form to another \cite{Stickel1996}. $T_B$ is typically
about $20\un{K}$ below $T_A$ and corresponds to the upper
temperature limit where dynamic scaling or Vogel-Fulcher
quantitatively describe the temperature dependence of dynamics.
%---------- 1st ref to TaPLOT -
%------------------------------

\begin{figure}[t]
\centerline{\includegraphics[width=3.1in,keepaspectratio=true]{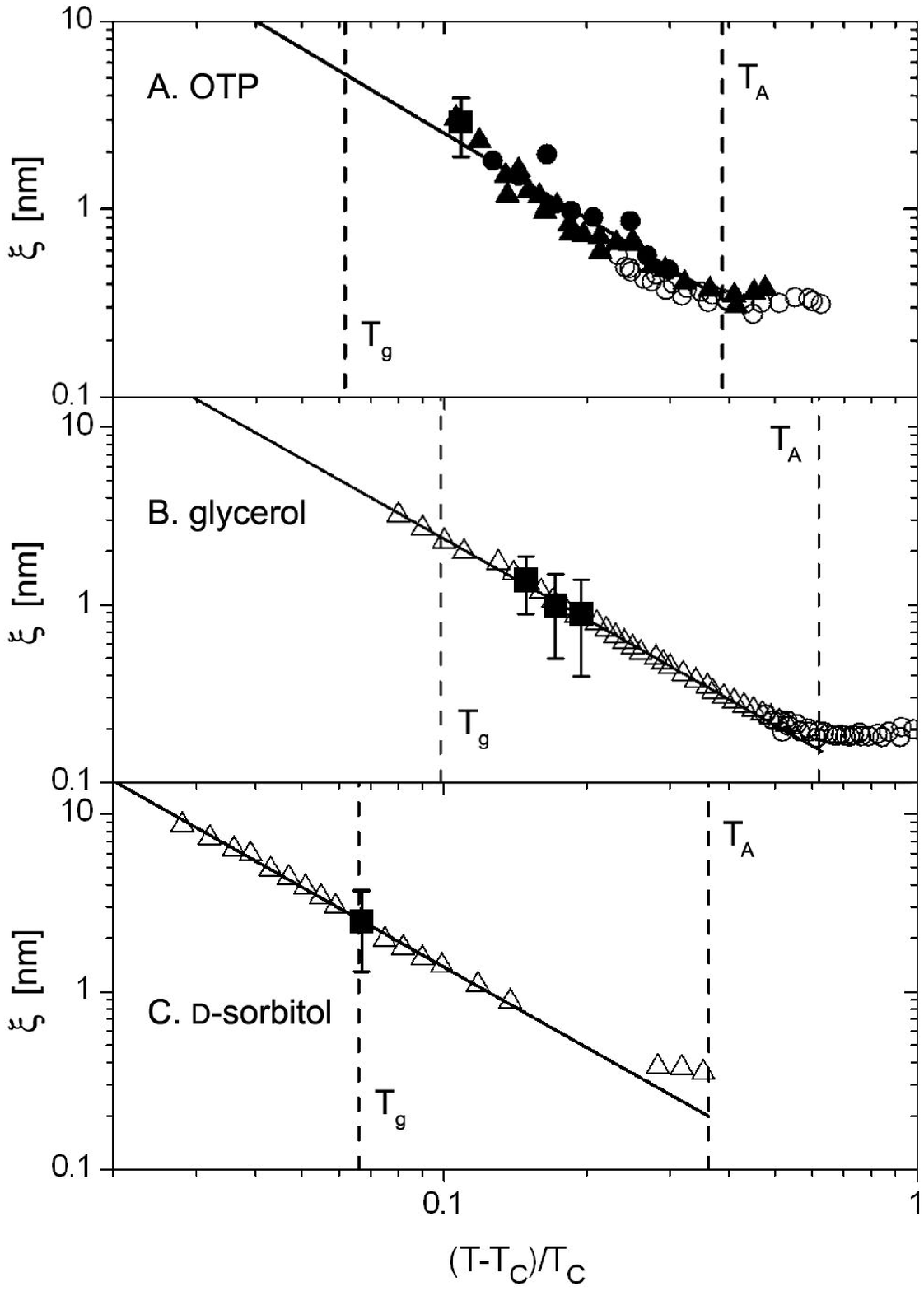}}
\caption{\label{xiPlot} The temperature dependence of the
cooperative length of OTP, glycerol and {\scriptsize\sf D}-sorbitol.
Filled symbols denote absolute measures of $\xi$ from 4-D NMR
($\blacksquare$ with error bars \cite{Qiu2002}) and calculations
using probe dynamics ($\bullet$ anthracene \cite{Cicerone1995A};
$\blacktriangle$ tetracene \cite{Cicerone1995A}). Open symbols
denote measures of the temperature dependence of $\xi$ from
self-diffusion ($\circ$ \cite{Fujara1992, Chang1997}) and dielectric
measurements ($\triangle$ \cite{Stickel1995, Nozaki1998}) which have
been vertically shifted into agreement with the absolute measures of
$\xi$. Dashed lines denote $T_A$ and $T_g$ (values in
Table~\ref{Table}). The solid line is the slope of $-3/2$ expected
by dynamic scaling (Eq.~\ref{ksi}).}
\end{figure}

The Stokes-Einstein relation expects the rotational relaxation time
$\langle\tau_r\rangle$ and translational diffusion coefficient $D_t$
of probe molecules are coupled so that their product is independent
of temperature. Above $T_A$ the probes diffuse a distance of order
their own size in the time it takes for the probe to rotate.
However, at $T_A$ these two dynamics \emph{decouple}, a fact which
can be used to establish a length scale. To understand this,
consider a bimodal distribution of $\phi_f$ fast and
$\phi_s=1-\phi_f$ slow particles. The average rotational time is
dominated by the slow particles
\begin{equation}
\langle\tau_r\rangle=\phi_s\tau_s+\phi_f\tau_f\cong\phi_s\tau_s
\end{equation}
since $\phi_s\gg\phi_f$ and $\tau_s\gg\tau_f$. The diffusion
coefficient is the sum of fast and slow contributions
\begin{equation}
D_t=\phi_sD_s+\phi_fD_f=\phi_s\xi^2_s/6\tau_s+\phi_f\xi^2_f/6\tau_f
\end{equation}
where $\xi_s^2\equiv6D_s\tau_s$ and $\xi_f^2\equiv6D_f\tau_f$. Since
$\phi_s\approx1$, the product
$6D_t\langle\tau_r\rangle\cong\phi^2_s\xi^2_s+\phi_f\phi_s\xi^2_f\tau_s/\tau_f$
can be used to define a length scale,
\begin{equation}\label{probe}
\xi\cong\xi_s=\sqrt{6D_t\langle\tau_r\rangle} \textrm{.}%
\end{equation}
Fig. \ref{xiPlot}.A shows the temperature dependence of this length
scale for OTP and it \emph{quantitatively} agrees with 4-D NMR,
which is known to target the slow contribution $\xi_s$.
%---------- 1st ref to xiPLOT -
%------------------------------
Hence, $\phi_s^2\xi_s^2\gg\phi_f\phi_s\xi_f^2\tau_s/\tau_f$,
justifying Eq.~\ref{probe}.

In the calculation of $\xi$, $\langle\tau_r\rangle$ was interpolated
\cite{Bainbridge1997} using the temperature dependence of viscosity
from various sources \cite{Plazek1994, Cukierman1973, Laughlin1972}.
In OTP, $\xi$ of anthracene (filled circles in Fig.~\ref{xiPlot}.A)
shows a strong temperature dependence over the entire experimental
temperature range because all $T<T_A$, whereas the larger tetracene
probe (filled triangles in Fig.~\ref{xiPlot}.A) is clearly
temperature dependent below $T_A$ and independent of temperature
above $T_A$.

$T_A$ is the temperature below which $\xi$ becomes temperature
dependent. Cooperativity causes the breakdown of the Stokes-Einstein
relation, as is made evident when $D_t\eta/T$ becomes temperature
dependent at temperatures below $T_A$ (Fig.~\ref{TaPLOT}).

Ideally a probe molecule would have the same shape, size, polarity
and properties as the liquid matrix in which it was inserted.
$^1$H-NMR can provide measurements of translational and rotational
\emph{self}-diffusion coefficients \cite{Fujara1992}. This allows
for direct measurement of $\xi$ using Eq.~\ref{probe}. Although
measurements of $\langle\tau_r\rangle$ exist \cite{Fujara1992}, they
are rare. In the case where measurements of $\langle\tau_r\rangle$
are lacking, we assume that $\eta/T$ properly describes the
temperature dependence of $\langle\tau_r\rangle$ \cite{Fujara1992}.
This assumption allows for measurements of $T_A$ that stand in
agreement with the other measurements of $T_A$, but with the
limitation of not being able to give an absolute measure of $\xi$.
Each of the glass-formers in Figure~\ref{TaPLOT} show that $T_A$
measured using dielectric and viscosity data occurs at the same
temperature as the breakdown in the Stokes-Einstein relation for
self-diffusion standing in agreement with the small probe data.

In Fig.~\ref{xiPlot}.A, the $\xi$ of OTP from self-diffusion
measurements are plotted along with the absolute measurements of
$\xi$ provided by 4-D NMR ($\xi (252K) = 2.9\pm 1\un{nm}$,
\cite{Reinsberg2002}). While each technique measures $\xi$
differently, they \emph{all stand in quantitative agreement}. The
$\xi$ from self-diffusion data also shows qualitative agreement with
the molecular probe data. While $T_A$ has been determined by
applying dynamic scaling to viscosity and dielectric measurements,
results from probe molecules and self-diffusion data are in
agreement since $\xi$ becomes temperature-independent above $T_A$.

Equation \ref{ksi} can be used to estimate $\xi$ from $\eta$ or
$\tau_\alpha$, since $T_C$, $E_\alpha$ and $E_\eta$ have been
previously established by fitting Eq.~\ref{ksi} to data.
\begin{equation}\label{xiDiel}
\xi \sim \tau_\alpha^{1/6}e^{-E_\alpha/6k_BT} \qquad T_C<T<T_A
\end{equation}
In Figs.~\ref{xiPlot}.B and \ref{xiPlot}.C, dielectric data have
been used to plot $\xi$ of glycerol and {\scriptsize\sf D}-sorbitol
along with measurements from 4-D NMR and in the case of glycerol,
self-diffusion. $\xi$ from dielectric and self-diffusion data were
vertically shifted into agreement with $\xi$ from the absolute
measurements that have been provided by 4-D NMR. The resulting data
obey the slope of $-3/2$ expected by dynamic scaling
(Eq.~\ref{ksi}). The $\xi$ plotted in Figures~\ref{xiPlot}.B and
\ref{xiPlot}.C shows a clear temperature dependence below $T_A$.

$\xi(T_A)$ and $\xi(T_g)$ are determined from Fig.~\ref{xiPlot} and
listed in Table~\ref{Table}. Uncertainty in the measurement of $\xi$
by 4-D NMR is extrapolated to $T_A$ and $T_g$ in cases where 4-D NMR
provides the only absolute measure of $\xi$ used in the
determination of $\xi$, as is the case with glycerol and
{\scriptsize\sf D}-sorbitol. The van der Waals sphere radii
$r_{vdW}$ of each of the glasses was calculated from atomic radii
using the procedures of Edward \cite{Edward1970} and are included in
Table \ref{Table}. The $r_{vdW}$ for {\scriptsize\sf D}-sorbitol,
OTP and glycerol are all within the calculated range of $\xi(T_A)$,
suggesting that the magnitudes of cooperative size calculated herein
are reasonable.

Four robust experimental methods for determining the caging
temperature have been identified. Of these methods, estimation of
the length scale for cooperative motion is best done from
measurements of probe diffusion and rotation. However, this has only
been done for a select few glass-formers. Far more convenient
techniques of dielectric spectroscopy and rheology can determine the
\emph{temperature dependence} of the length scale.

Despite the fact that the cooperative volume is fractal instead of
space-filling, the essential features of the Adam-Gibbs model
\cite{Adam1965} are correct. The cooperative size does indeed grow
rapidly as temperature is lowered below the caging temperature. Over
a temperature range extending from $T_A$ to $T_g$, $\xi$ has been
observed to increase by an order of magnitude from the van der Waals
radius of each molecule, making the insight of Adam and Gibbs
particularly noteworthy.

We thank the National Science Foundation (DMR-9977928 and
DMR-0422079) for funding.

\end{document}